# DCBA: Detection of Collaborative Black-Hole Attacks in Connected Dominated Set using Baiting Process

**Dr. Lata B T[1] and Dr. Venugopal K R[2]**

[1]Associate Professor, CSE Dept, UVCE, Bengaluru, India

[2]Former Vice Chancellor, BU, Bengaluru, India

## Abstract

Mobile Ad-hoc Network (MANET) is temporary and dynamic network topology, wherein nodes are mobile in nature and distributed randomly in a network area. In MANET, nodes cooperate with each other to operate and forward data through multihop communication between source and destination. MANET is exposed to different types of attacks due to absence of central administration. However, some nodes decline to cooperate, misbehaves and appears to be malicious affecting network functionality and connectivity. Providing security and identifying malicious node has become one of the challenging research topics in MANET. Black-hole attack is considered to be most popular attack that degrades the overall network performance. Black-hole node falsely advertises the shortest path to destination intentionally to disrupt the network communication resulting in packet drop. In collaborative black-hole attacks, multiple black-hole nodes cooperate and launch attacks in order to degrade network reliability. In this article we propose a Lightweight technique to detect and isolate Collaborative Black-Hole attacks (LW-CBH) by enhancing existing AODV routing protocol. In this scheme a timer based baiting process and reverse tracing setup is used to detect malicious node through control status message in MAC layer which are Reply Sequence (R-SEQ) and Code Sequence (C-SEQ) message of connected dominated set of nodes. However existing AODV routing protocol fails to detect malicious node during dynamic topology changing in MANET. Simulation of proposed technique is performed using discrete event simulator tool NS-2.35. The simulation results are evaluated for throughput, packet delivery ratio, average end-to-end delay and normalized routing overhead.

## I Introduction

Expanding wireless network, Mobile Ad-hoc Networks (MANETs) is an ideal solution which enables nodes to communicate without any infrastructure. Self-configuration, dynamic topology, smaller, powerful, adaptive and less expensive device features make MANETs a prominent reason for growing popularity in wireless network [1,2]. MANET's applications range from military operation, mission critical networks, crisis management and automotive communications [3, 4].

Dynamic nature of mobile nodes leads to frequent change in network topology. Routing protocols in MANET's are designed to adapt dynamic network topology changes for reliable data transfer [5]. Nodes are equipped with battery and energy is one of the most important factors for network connectivity. Energy consumption of nodes should be limited to extend node lifetime. Nodes are linked with wireless signal, link quality between two nodes should be stable for data delivery. Due to node mobility link failure occur frequently which leads to channel fading, packet drop and interference [6]. Due to unstructured network MANETs is vulnerable to different types of threats and attacks [7]. Since nodes are exposed to open environment and wireless links, data can be altered and dropped intentionally by attackers. Some malicious nodes enter and try to disturb the network by misbehaving, manipulating and dropping data packets. This type of phenomena is called a Denial of Service (DoS) attack [8]. In the event of DoS attack the attacker node tries to drain the nodes energy and degrades the overall network performance. Black-hole attack is considered to be one the protuberant DoS attack. Due to open medium black-hole node can easily enter into network and advertises itself of having shortest route to destination during route discovery process. Normal nodes get compromised to black-hole node by calming false information and starts dropping the packets intentionally. More than one black-hole node collaborate with other black-hole in order to degrade the network performance is known as collaborative black-hole attacks [9].

In this paper we propose lightweight technique to defend collaborative black-hole attack through timer based cooperative baiting process using false id to bait black-hole node that responds to bait request which does not sent by the legitimate nodes at that particular time instant. In reverse tracing each nodes status is checked before accessing channel to connect network layer by issuing control status message from MAC layer to verify information of code sequence, reply sequence and energy level to all neighbour nodes. Legitimate nodes reply without manipulating any information, whereas malicious replies with forged information to prevent its identity. Detection of malicious node becomes easier with the fake information and mismatch in the control status message. If control status message match, then nodes are allowed to connect to network layer for routing data otherwise node will be discarded and malicious node information is broadcasted to entire nodes in the network to block communication.

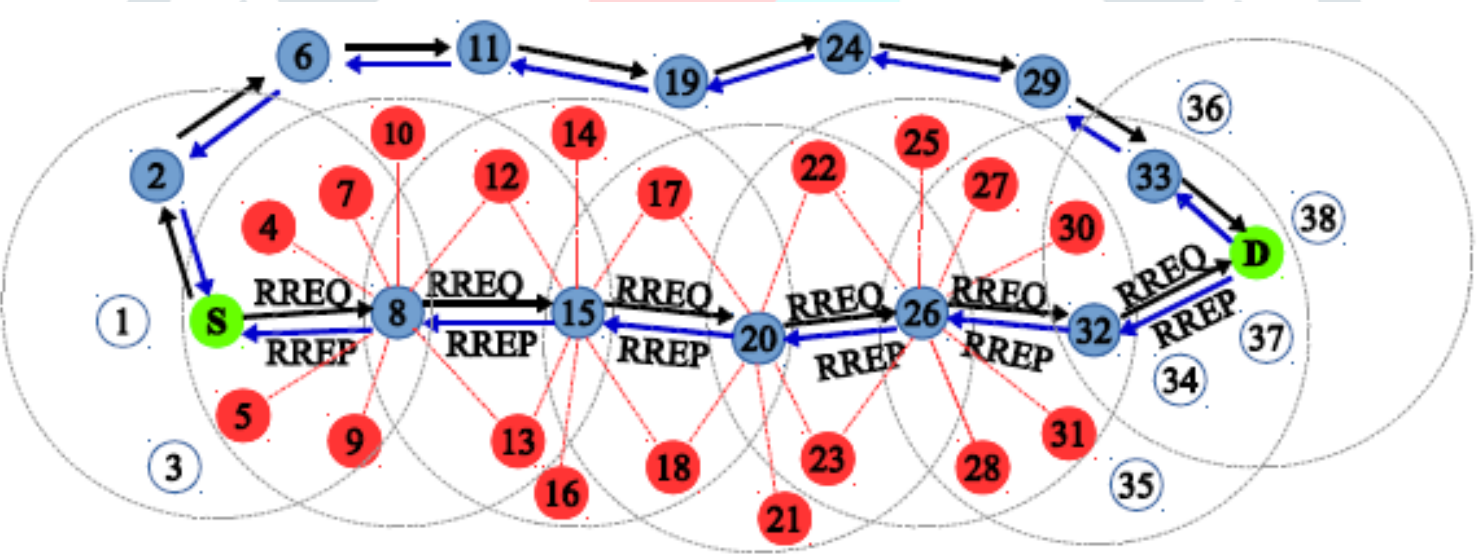

**Figure 1: MANETs Routing Process**

## II Problem Statement and Motivation

Securing MANET's communication is essential to prevent from malicious node that causes harm to network. MANET's are highly vulnerable to different types of attacks due to lack of infrastructure, open environment and dynamic topology changes. Black-hole attack is considered to be popular DoS attack that harm the network and degrade the overall network performance [10]. Ad-hoc On-demand Distance Vector (AODV) routing protocols finds the best and optimal path from source to destination and it is proven that AODV routing protocol delivers highest packet ratio [11-13]. However, AODV routing protocol fails to provide secure routing between source and destination. Many research works have focused on providing secure routing for MANET's to deal with detection and prevention of individual black-hole node. When network has multiple black-hole nodes, some approaches fail to detect and network becomes weaker and ineffective. This motivates us to propose a technique to detect and isolate the collaborative black-hole node and provide secure routing between source and destination.

The rest of this paper is organized as follows. Section II describes problem statement and motivation. Section III describes the related works done. Section IV describes the proposed network model, adversary attack model. In section V describes proposed LW-CBH scheme. In section VI the performance analysis and relative simulation are conducted. Finally, we draw the conclusion on the proposed scheme in section VII.

## III Related Works

In [14] author proposed MBDP-AODV protocol to detect black-hole node based on dynamic sequence number threshold. Mean dynamic threshold values are calculated by source node through standard deviation of destination sequence number. MBDP-AODV protocol includes two packets SUSPECT packet, used to detect lair node and ALERT packet to broadcast the lair node information. It is found that MBDP-AODV performs well in packet delivery, throughput and mitigates black-hole attack under varying node density but faces high routing overhead. In [15] author proposed detection mechanism for black-hole and gray hole attack by implementing two version of invincible AODV routing protocol during route discovery process. First version includes of checking node to node frame check sequence tracking and second signed frame check sequence tracing. This authentication mechanism provides secure reliable data transmission to all legitimate nodes showing significant improvement in end-to-end delay and packet delivery ratio. However, this mechanism has higher computational complexity and slow working process.

In [16] author proposed secure routing to prevent black-hole attack by setting validity bit in control message (RREP). Malicious node is unaware of validity bit that has to be sent to source node while sending RREP message. Source node upon receiving RREP message, source node check for validity bit if it is set to 1, then source connects to that path else source assumes that RREP if from black-hole node and discards it. Limitation of this scheme is unrealistic and any smart black-hole node can analyse and notice validity bit.

In [17] proposed trust based frame work to detect and prevent black-hole node in MANET's. In this framework node behaviour is observed using different trust metrics and analysed in terms of communication and mobility behaviour. Zone routing protocol (ZRP) is used to retain data privacy of node. This framework shows overall network performance in terms of throughput, trust level and achieves QoS. However, this framework has higher routing overhead.

In [18] author proposed Cooperative Bait Detection Scheme (CBDS) to detect multiple black-hole nodes. In CBDS source node sends a bait request to one of its randomly selected neighbours. List of suspicious nodes are created from RREP message for bait request to detect black-hole node. Finally black-hole is defended by determining the PDR threshold value set by the source node. Limitation of CBDS is that nodes enter into promiscuous mode. In [19] author proposed model to mitigate black-hole attack for military perspective. In this model fake request is flooded to network, node responding to this fake request is considered as suspicious node. By checking the suspicious node activity of forwarding packets to destination by legitimate neighbour node can detect black-hole node. This model leads to congestion due to flooding of fake request message making a drawback.

## IV Network Model

Network consists of mobile nodes randomly deployed in an area, nodes move randomly in two-dimension using random way point (RWP) model. Localization algorithm provides information of mobile nodes speed, current location and direction at each time $\Delta t$ [20]. Nodes movement in network is unpredictable, random waypoint model (RWP) periodically estimates nodes mobility. Nodes are homogeneous and assigned with same amount of initial energy.

### Adversary Attack Model

We consider black-hole attack, it is an active type of attack where intruder node drops full packet and claims fake shortest route to any destination even if it does not have any route. When source node has packets to send to destination, source node broadcast route request packet (RREQ). Neighbour node within the transmission range replies to RREQ by sending reply packet (RREP). Normal node upon receiving RREP trusts any node reply message and black-hole node takes advantage of sending false RREP and claims that it has shortest path to any node. Since there is no technique to verify the RREP message is from normal or black-hole node. Source node initiates packet forwarding to black-hole node hoping to deliver it to destination, intentionally black-hole node starts dropping the entire packet it received. When one attacker node is active is a single black-hole, more than one attacker node cooperate with other is a collaborative black-hole attack. In this paper we consider collaborative black-hole attack.

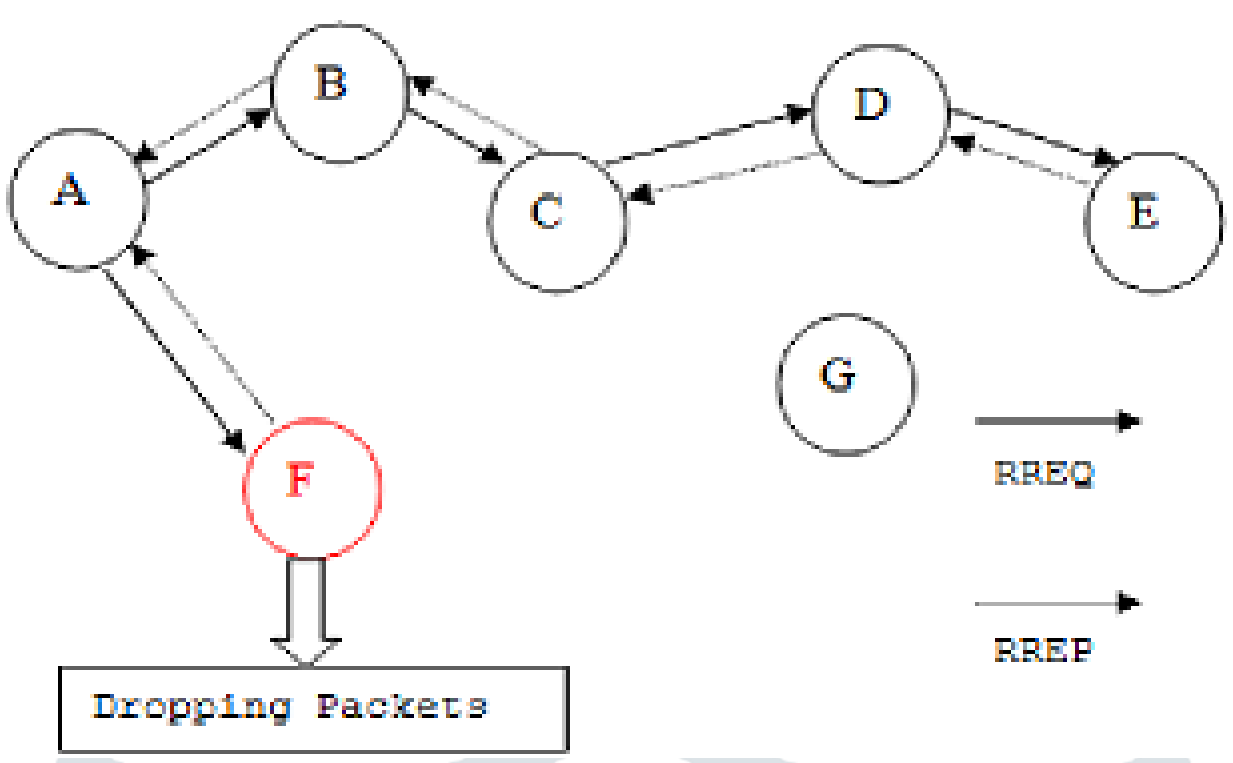

**Figure 2: Black-hole attack scenario**

## V Proposed LW-CBH Scheme

Proposed scheme is to detect and isolate collaborative black-hole attack, in baiting process the false id is used to bait and detect black-hole nodes in network. During baiting process timer value is set randomly to broadcast the bait request through false id. Upon receiving the bait request black-hole nodes replies to source node and claims shortest path even if it does not exist. Source node immediately considers the node responding to bait request with fake id as black-hole node and adds it to the malicious node list. To avoid network congestion, the bait request is added with TTL (Time-To-Live) field and it is set to one. During route discovery AODV routing protocol broadcasts RREQ packets from source to destination to communicate with another node in the network. For secure data routing we introduce two control status messages code sequence (C-SEQ) and reply sequence (R-SEQ) in MAC layer while accessing channel. Code sequence message is sent by intermediate nodes to its neighbours. Intern each neighbour replies by sending reply sequence message to intermediate node. If the control status message matches from the neighbours then the connection to the network layer by accessing channel will be allowed by intermediate node else it will be discarded and information is sent to all nodes as malicious. Each node sends reply sequence message to source node for reverse trace path of code sequence. However malicious node replies with fake id to source thus control status gets mismatched.

---

**Algorithm to detect malicious node using LW-CBH:**

---

**Input: Baiting to detect malicious node**
**Output: Isolate malicious node**
**RREQ: Request message, C-SEQ: Code sequence message, R-SEQ: Reply sequence message, Dest_SEQ_ID : Destination sequence id, SEQ_ID : Sequence id**

**At source node**

- if curr_time = = baiting_time then
- do bait_request
- generate bait_request (false_id) by setting random ID
- set TTL to bait_request as 1
- broadcast bait_request

- reset baiting_time to random time
- end_ if
- for each bait_request (false_id) received reply do
- add false_id nodes to black-hole list
- end_for

**For secure routing**

- Initialise process
- Source node initiates route discovery process using RREQ, status C-SEQ and R-SEQ
- Store RREP, Dest_SEQ_ID and Neighbour_ID in C-SEQ list
- Identify malicious node by retrieving C-SEQ list
- if Dest_SEQ_ID is greater than Sending SEQ_ID discard from C-SEQ list
- Mark node as malicious = Node_ID (false_id)
- Neighbour node selection process
- Sort C-SEQ entries and select Neighbours having valid RREP
- Continue default process
- Allow valid node to access channel using AODV-MAC routing protocol

**Energy consumption**

In proposed scheme every node is assigned with same initial energy, node consumes energy in the process of transmitting, receiving and mobility. Energy is supplied from the battery and it is limited. Consumption of energy should be optimal to extend network lifetime and energy of nodes should be balanced for reliable data transfer. The energy consumption of node for various processes in the network is given as:

$$N_{e(t)} = N_e(t-1) - \left(N_{e_{tx}} \times s_{t-1}\right) - \left(N_{e_{rx}} \times s_{t-1}\right)$$

where $N_{e(t)}$ represent residual energy at time $t$, $s_{t-1}$ is the data transmitted at time $t-1$. Energy required to transmit is given as $N_{e_{tx}}$ and for receiving data bytes at reception is given as $N_{e_{rx}}$

An residual energy field is added in the RREQ message in AODV routing protocol, after selecting neighbours from the secure routing algorithm every intermediate node adds its residual energy for reliable path selection. Average path energy at destination is calculated as:

$$N_{avg} = \frac{N_Eg}{nh}$$

$N_{avg}$ represents average residual energy of path, total amount of energy required for path is given as $N_Eg$ and $nh$ is number of hop count. The destination node selects the nodes which has higher residual energy and stores into routing table.

**Energy consumption Algorithm**

- if message = RREQ and residual energy > threshold value
- {
- $N_{avg} = N_{Eg} + residual\ energy$
- route packets
- }
- else
- discard

## VI Results and Evaluation

### Simulation parameters

Performance of proposed LW-CBH scheme is compared with DDBG [21] scheme using IDS detection of black-hole and grey-hole attacks. Performance analysis is evaluated for collaborative black-holes in the network. The network scenario includes varying the nodes node size of 20, 30 and 40. Simulation time is set to 150 seconds by keeping the node mobility speed constant to 25m/s. Simulation is carried out on network simulator tool (NS2) [22]. Constant bit rate (CBR) traffic is utilised, the nodes are randomly deployed in an network area of 1000×1000 mts. Nodes are mobile and topology change occurs frequently as nodes move in different velocities and speed. The node transmission range is set to 250 m and nodes are assigned with initial energy of 100 Joules and threshold energy value is set to 50 Joules. Simulation parameters used is presented in Table no 1:

| Simulation Parameters | Value |
|---|---|
| Network terrain | 1000x1000 |
| No of nodes | 20,30,40 |
| Radio | 802.11 |
| Transmission range | 250mts |
| Traffic | CBR |
| Simulation time | 150 sec |
| Initial energy | 100J |
| Packet size | 512bytes |
| Malicious nodes | 2 |

**Table No 1: Simulation parameters**

### Performance Evaluation

**Packet Delivery Ratio:** is the ratio of transmitted packet from the source to the received packets at the receiver to destination [23]. This metric indicates a routing protocol's reliability in its transmission of data packets from source to destination. The higher the ratio, the better the routing protocol efficiency will be.

$$PDR = \frac{number\ of\ received\ packets}{number\ of\ transmitted\ packets} * 100$$

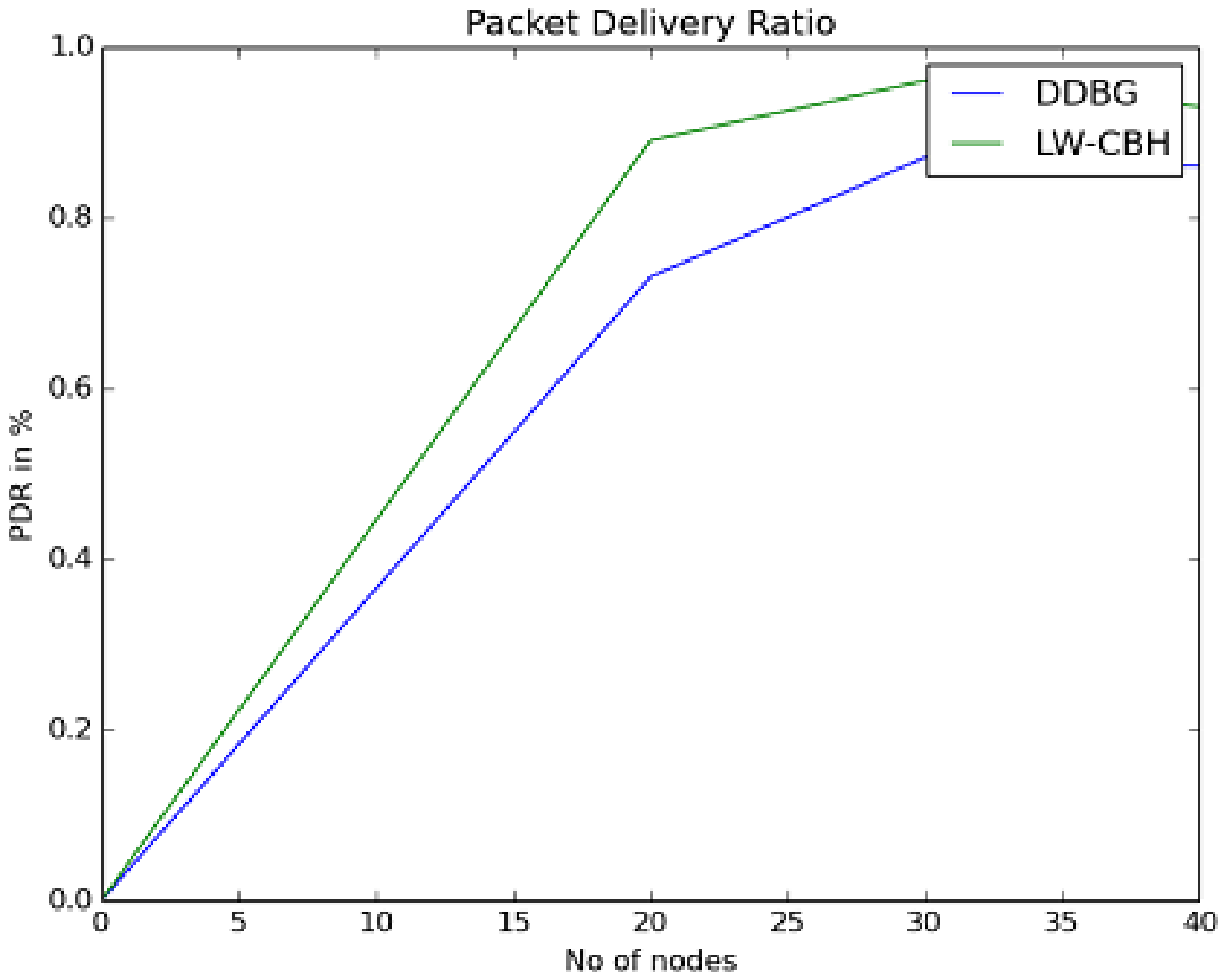

**Figure 3: PDR vs No of Nodes**

As shown in Figure 3 by varying nodes the result of LW-CBH shows better performance than DDBG scheme in delivering highest packet ratio in presence of black-hole nodes. It is observed that when node size increases the black-hole nodes collaborate and tries to send more fake id's and disturb the network. However our proposed scheme detects malicious node through baiting process and verifying the control status message before routing the actual data packets.

**Throughput:** is defined as the number of bits obtained successfully at destination. It is expressed as kilobits per second (Kbps) [24]. Routing protocols efficiency is measured by receiving data packets at destination. Throughput is calculated as:

$$Throughput = (total\, received\, bytes\ * 8/\ time) * 1000\ Kbps$$

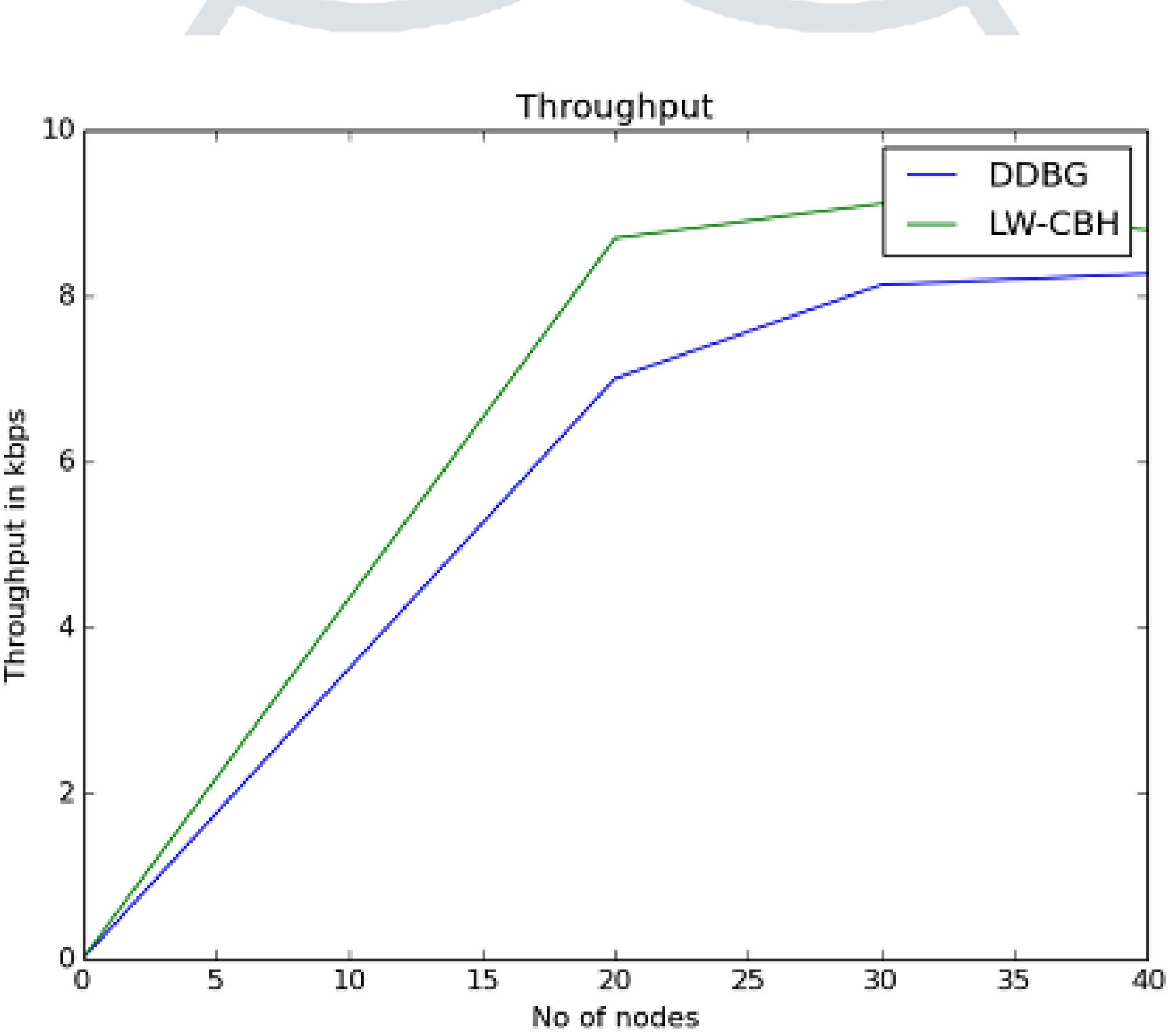

**Figure 4: Throughput vs No of nodes**

Figure 4 shows the network throughput graph when data is transferred between two nodes. As shown in the throughput graph the proposed scheme achieves high throughput compared to DDBG scheme. Increase in number of nodes makes possible to discover many routes between source and destination. To validate different paths, in the presence of malicious node LW-CBH uses timer based baiting scheme to detect malicious nodes claiming shortest path to destination even if it does not exist. Due to mobility of nodes, frequent link failure occurs and topology changes this make node to expose to malicious attacks. LW-CBH scheme discovers safer routes by sending RREQ and checks the control status messages C-SEQ and R-SEQ and allows accessing channel for reliable data routing over the different paths.

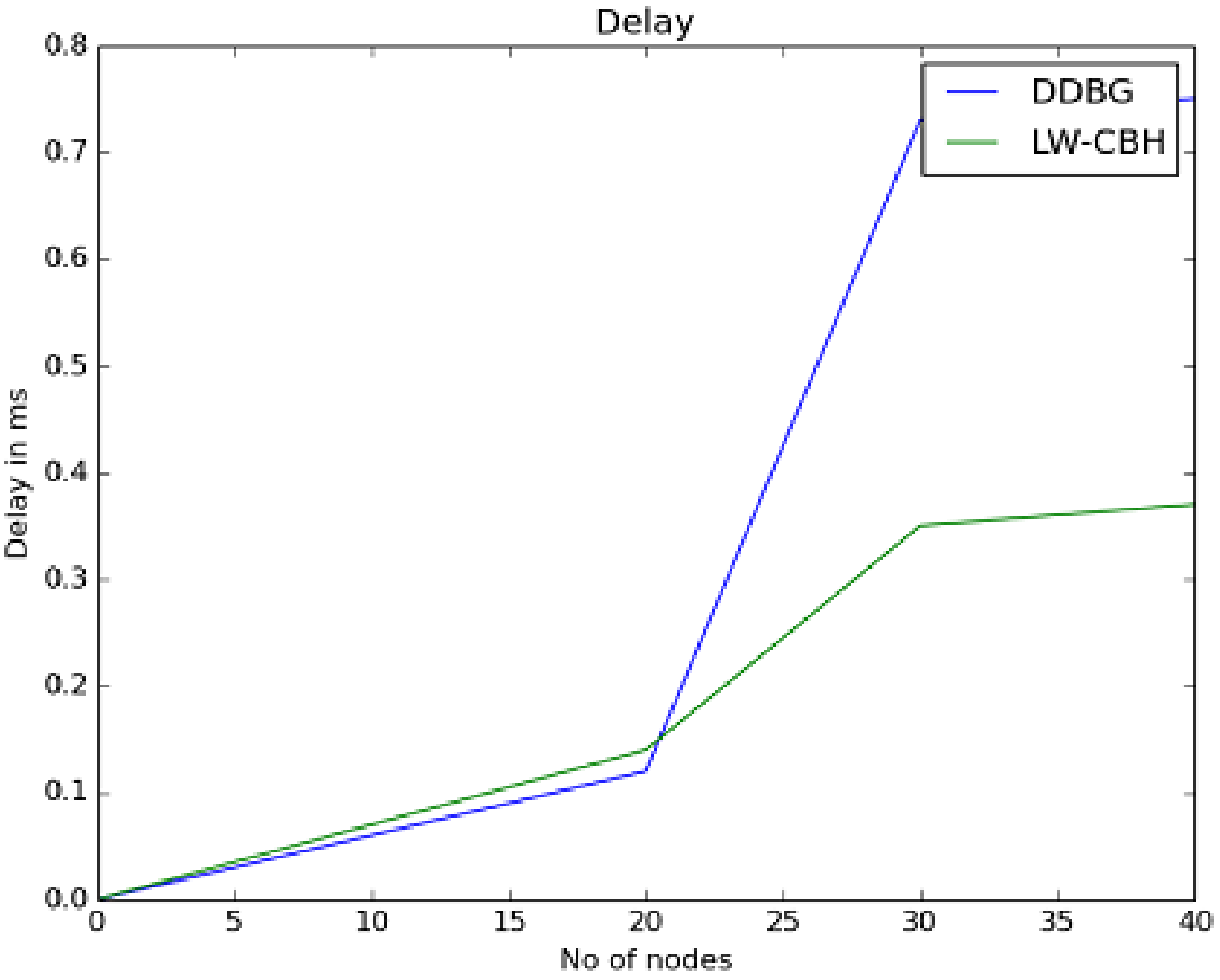

**Figure 5: Delay vs No of Nodes**

**End-to-end Delay:** is defined as delay time required receiving the transmitted data packet and retransmitting to other nodes across the network [25].

$$E2E = \sum T_1 - T_2 / N$$

where $T_1$ first data packet arrived time, $T_2$ is time first packet sent by source and $N$ is total number of packets sent. Figure 5 shows the delay result of DDBG and proposed LW-CBH scheme by means of the time required to deliver the packets to destination. The results shows delay of DDBG scheme increases due to retransmission happened after detecting the malicious node by computing the IDS functions. In our proposed LW-CBH scheme minimizes the retransmission of packets due to AODV-MAC layer which allows node to carry data through channels only if the conditions are satisfied such that the nodes in the path are safer to carry data to destination. As the node number increases the computation of average route delay increases due to link breakages and finding alternate routes by sending RREQ message. It can be seen that proposed scheme performs better than DDBG scheme.

**Routing Overhead**

The normalized overhead of routing (NRO) [26] is determined according to equation below:

$$Routing\ Overhead = \frac{N_1}{N_2}$$

such that the total sum of control packets associated with The data packets are divided by the total number of data packets that the destination node receives. where $N_1$are number of routing packets sent and forwarded, $N_2$ is number of data packets received.

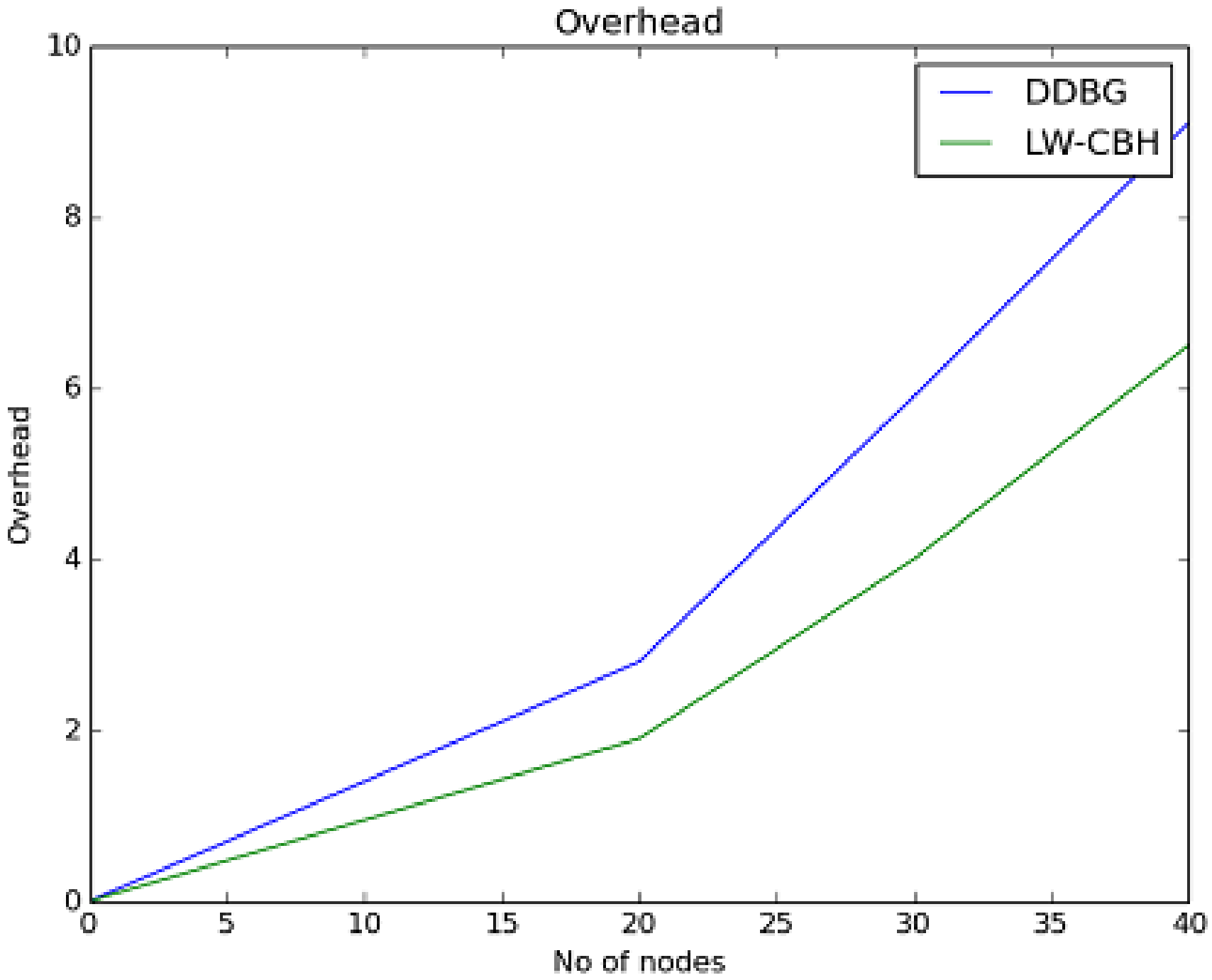

**Figure 6: Normalized routing vs No of nodes**

Figure 6 shows the normalized routing overhead. It has been seen that DDBG scheme has high routing overhead due to computing IDS and queries functions every time during routing process which increases nodes computing functions. In our proposed LW-CBH scheme the computing function are light weighted and detects the malicious node by sending false bait request where the malicious node replies to false request. Therefore LW-CBH scheme does not burdens nodes with high routing computation. Routes are computed only after detecting malicious nodes which replies to false request. Actual route discovery process starts only after RREQ, C-SEQ and R-SEQ matches.

## Conclusion

Different kind of attacks disturbs the network operation in MANETs which is a major concern. Detection of malicious activities in the network leads to decrease in overall network performance. It is essential task to detect and isolate malicious node for secure data routing. It is considered that the black-hole attack is one of the most severe attacks that impact MANET's operations. The detection and isolation of black-hole in the network is considered an important role to avoid threats which leads to network breakdown. Black-hole node can be of single or cooperative black-hole nodes. In order to detect and isolate malicious nodes in the network we have proposed a light weight scheme to detect collaborative black-hole attack (LW-CBH) in MANET's. This scheme employs timer based cooperative baiting process using false id to bait black-hole node that responds to bait request which does not sent by the legitimate nodes at that particular time instant. For secure routing the control status message code sequence (C-SEQ) and reply sequence (R-SEQ) are used in AODV-MAC layer which matches the control messages to allow channel access otherwise it discards the node accessing the channel. Simulation of LW-CBH is compared with DDBG scheme, simulation results prove that our proposed scheme detects malicious node successfully and differentiate between fake replies and legitimate response from nodes. An experimental outcome of LW-CBH is an effective and prominent approach to detect collaborative black-hole attacks. As a future work and effective trust model to detect different kind of attacks may be designed to enhance the network parameters.